# QoS Based Capacity Enhancement for WCDMA Network with Coding Scheme

K.AYYAPPAN[1] and R. KUMAR[2]

[1]Rajiv Gandhi College of Engineering and Technology, India.
*aaa_rgcet@ yahoo.co.in*
[2]SRM University, Chennai, India
*rkumar68@gmail.com*

## ABSTRACT

*The wide-band code division multiple access (WCDMA) based 3G and beyond cellular mobile wireless networks are expected to provide a diverse range of multimedia services to mobile users with guaranteed quality of service (QoS). To serve diverse quality of service requirements of these networks it necessitates new radio resource management strategies for effective utilization of network resources with coding schemes. Call admission control (CAC) is a significant component in wireless networks to guarantee quality of service requirements and also to enhance the network resilience. In this paper capacity enhancement for WCDMA network with convolutional coding scheme is discussed and compared with block code and without coding scheme to achieve a better balance between resource utilization and quality of service provisioning. The model of this network is valid for the real-time (RT) and non-real-time (NRT) services having different data rate. Simulation results demonstrate the effectiveness of the network using convolutional code in terms of capacity enhancement and QoS of the voice and video services.*

## KEYWORDS

## 1. INTRODUCTION

Third generation wireless communication systems are designed for multimedia services such as audio, video and data communication with enhanced high data rates. This will create new opportunities not only for manufacturers and operators, but also for the service providers of such applications using these networks. In the standardization forums, wideband code division multiple access (WCDMA) technology has emerged as the most widely adopted third generation air interface network [1-4]. Its specification has been created in 3rd generation partnership project (3GPP), which is the joint standardization project of the standardization bodies from Europe, Japan, Korea, the USA and China. Within 3GPP, WCDMA is called universal terrestrial radio access (UTRA) to cover both frequency division duplex (FDD) and time division duplex operation (TDD).

The universal mobile telecommunication system (UMTS), a WCDMA based network, is required to support a wide range of applications each with its own specific QoS requirements. There are four distinct QoS classes of service namely, conversational, streaming, interactive and background. Each class has its own QoS specifications such as delay and bit error rate (BER). One of the main challenges in 3G and beyond wireless networks is to guarantee QoS requirements while taking into account radio resource limitations. Call admission control is a technique to manage radio resources for optimizing the overall network performance. CAC is one of the radio resource management (RRM) process of making a decision for a new call admission taking into account the amount of available resource and users QoS requirements [5].

In call admission control strategies, a user originates a call to the network requesting a desired QoS; the network must check two things before accepting the call request. First, network must make sure that it has sufficient bandwidth to allocate to the user. Second, it must determine if, after admitting the user, it can continue to provide the same QoS for all existing connections. Thus, before the network admits the new user, it should determine whether it can meet the QoS for all connections, old as well as new. The capacity of WCDMA cell is defined in terms of cell load where the load factor, is the instantaneous resource utilization bounded by the maximum cell capacity. Instantaneous values for the cell load range from 0 to 1.

Related work

Many researchers have been done on the capacity of WCDMA system. In [6] Alma Skopljak Ramovic determined the WCDMA capacity by processing gain, bit energy to noise density ratio, voice activity factor and the total interference. The interference is already included in noise power density and it comprises the multiple access interference (MAI), self interference and co-channel interference.

In [7] B Christer and V Johansson explained the optimum value of target Eb/No with respect to bit error rate (BER) to improve the system capacity. In a WCDMA packet data system if the target Eb/No is set too low then the system will suffer from many retransmissions. This will reduce the capacity of the network. However, the number of mobiles that a system can serve simultaneously is inversely proportional to the Eb/No target, so also a too high Eb/No target will result in a capacity decrease.

The parameters which impact on WCDMA capacity are based on type of services such as real-time and non-real-time services. Real-time applications need some guaranteed minimum transmission rate which requires reservation of system capacity. In [8] Hedge and E. Altman determined the capacity of multi-service WCDMA networks with variable grade of service (GoS). The inputs of multi-service cell dimensioning are the offered load, the required resource (effective bandwidth), the requirements on blocking for each service and the total resource available in the cell. For a given number of subscribers the blocking probabilities are different for different services because they share the same pool of resources. The more resources a service needs for one user the higher is the blocking probability [9, 10].

Traditional definition of capacity of networks are either related to the number of calls they can handle (pole capacity) or to the arrival rate that guarantees that the rejection rate (or outage) is below a given fraction. In this paper a quality of service based capacity enhancement for WCDMA networks is designed for different services by fixing thresholds to each service according to their utilization with convolutional coding scheme by selecting suitable Eb/No value with respect BER to maintain the quality of service.

This paper is organized as follows; Section 2 discusses the call admission control scheme for WCDMA wireless network Section 3 brings out the simulation results for downlink capacity calculation for different services with and without coding schemes. Section 4 concludes the paper.

## 2. CALL ADMISSION CONTROL SCHEME

Call admission control is one of the major tasks of radio resource management when a new connection is set up, the admission control will be used to guarantee that there are free radio resources. In addition, CAC determines which base station will have power control and must have sufficient bandwidth to support the new connection without dropping any of the existing ones. If this condition is not met, the new connection request will be rejected. This check is done whenever a user enters a new cell, either through a new call or handover call [11].

CAC for WCDMA

The two most commonly used call admission control schemes are wideband power based (WPB) scheme and throughput based (TB) scheme [12, 13]. The principle of this scheme is a new user is admitted when the new load factor after the new user's admission does not exceed the predefined threshold. The uplink and downlink directions are considered and only if the condition is met, the new connection request can be admitted.

The new load factor for the network uplink or downlink is the sum of the existing uplink or downlink load factor and the increase in load factor $\Delta L$.

$$\eta_{New} = \eta_{Old} + \Delta L \tag{1}$$

The computation of the uplink existing load factor and the downlink existing load factor is shown in (2) and (3) respectively

The computation of $\eta_{UL}$ is given by

$$\eta_{UL} = (1+i)\sum_{j=1}^{N} \frac{1}{1 + \frac{W}{(E_b/N_o)_j R_j v_j}} \tag{2}$$

where
    W is the chip rate
    $v_j$ is the voice activity factor of $j^{th}$ user
    R is the bit rate of $j^{th}$ user
    $E_b/N_o$ is the bit energy to noise density ratio of $j^{th}$ user
    $i$ is the total interference
    N is the number user

The computation of $\eta_{DL}$ is given by

$$\eta_{DL} = \sum_{j=1}^{N} R_j \frac{v_j (E_b/N_o)}{W}[(1-\alpha_{av}) + i] \tag{3}$$

$\alpha_{av}$ is the average orthogonality factor of the cell

The performance of utility function based CAC algorithms is analyzed in the downlink channel of WCDMA network. The downlink load factor is set to 0.7 to maintain the channel load without affecting the QoS. With in the load factor limit of the channel the load factor for different services are selected according to the service needs and QoS [14]. The new call request is admitted as per flow chart shown in Fig.1.

The new load factor for the downlink is the sum of the existing downlink load factor $\eta_{DL}$ and the increase in the load factor $\Delta L$. The new load factor cannot exceed a predefined threshold:

$$\eta_{DL} + \Delta L \leq \eta_{DLthershold} \tag{4}$$

$\eta_{DLthreshold}$ is the downlink load factor threshold value.

Initially the call is identified as voice call or data call and then it is checked for the resource availability as per load factor threshold value for individual service. The sum of new call load and existing connection load is less than the threshold value of individual services then the call is admitted otherwise it is rejected.

$$\eta_{DLvoice} + \Delta L \leq \eta_{DLvoicethreshold} \tag{5}$$

$$\eta_{DLdata} + \Delta L \leq \eta_{DLdatathreshold} \tag{6}$$

$\eta_{DLvoicethreshold}$ is the downlink load factor voice threshold value.

$\eta_{DLdatathreshold}$ is the downlink load factor data threshold value.

One major limitation of the fixed threshold schemes is that the reserved capacity for voice traffic classes may remain unutilized while video priority classes are being blocked. In the proposed scheme the unutilized resources are utilized by blocked users of the fixed threshold scheme [15]. This will improve the capacity of the network as well as reduces the call blocking probability of the network.

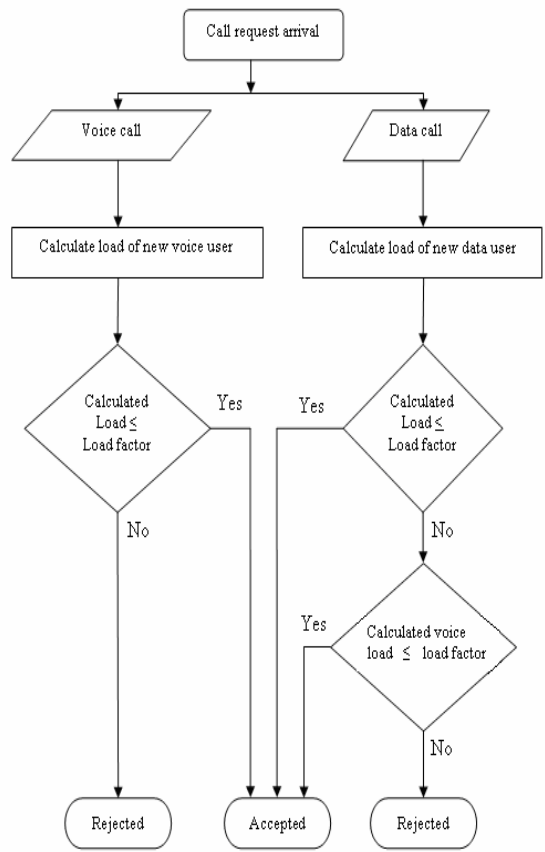

.

**Fig.1. Flow chart of utility based CAC scheme**

## 3. PERFORMANCE ANALYSIS

The objective of this simulation is to analyse the utility based CAC for different data rate services in WCDMA network. The simulation model is based on downlink load factor 0.7 with 70% voice users and 30% video users. The capacity of the network enhanced with utility function based call admission control with convolution code is compared to fixed load factor threshold setting in terms of capacity and number of blocked users. The simulation parameters which are used are shown in Table.1.

**Table1. Simulation parameters**

| S.No. | Parameter | Values |
|---|---|---|
| 1 | Chip rate (W) | 3.84 Mcps |
| 2 | Voice Bit rate (R) | 12.2 kbps |
| 3 | Video Bit rate (R) | 64 kbps |
| 4 | Voice Activity factor($v_j$) | 0.58 |
| 5 | Data Activity factor | 1 |
| 6 | Total interference (i) | 0.55 |
| 7 | Orthogonality factor ($\alpha$) | 0.9 |
| 8 | Voice Bit energy to noise density ratio (Eb/No) | 6.7dB, 5.7dB, 5dB |
| 9 | Video Bit energy to noise density ratio (Eb/No) | 9.6dB, 8.4dB, 6.5dB |

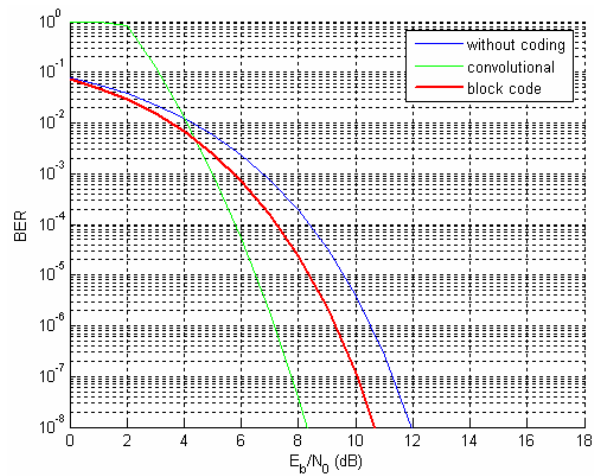

**Fig.2. Bit Error Rate to the Bit energy to noise density ratio**

The Fig.2 shows the bit error rate to the bit energy to noise density ratio (Eb/No) of BPSK modulation of WCDMA network for different coding scheme. To maintain QoS for voice service the BER is $10^{-3}$ and the corresponding Eb/No values are 6.7 dB for without coding, 5.7 dB for Block code and 5 dB for convolution code. Similarly for video service the BER is $10^{-5}$ and the corresponding Eb/No values are 9.6 dB for without coding, 8.4 dB for Block code and 6.5 dB for convolution code.

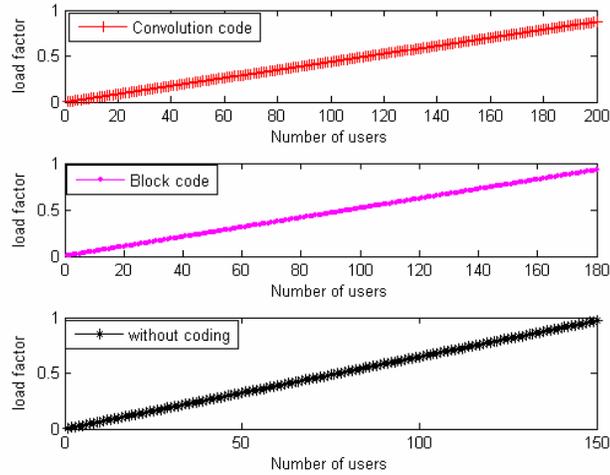

**Fig.3. Downlink load factor for voice service with coding**

The Fig.3 shows the number users admitted in the forward channel of WCDMA network for voice (12.2 kbps) service without coding scheme, with block code and convolution code. The number of users connected to the network is 109 voice users for without coding scheme, 137 users for block code and 160 users for convolution code for load factor threshold value of 0.7.

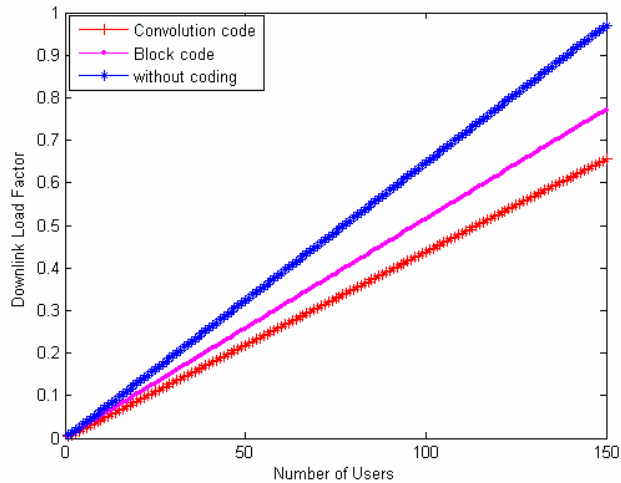

**Fig.4. Downlink load factor for voice service with coding**

The Fig.4 shows the capacity enhancement of WCDMA network for voice (12.2 kbps) service for different coding scheme. The number of users is 150 the corresponding downlink load factor is 0.9708 for without coding scheme, 0.7711 for block code and 0 .6563 for convolution code.

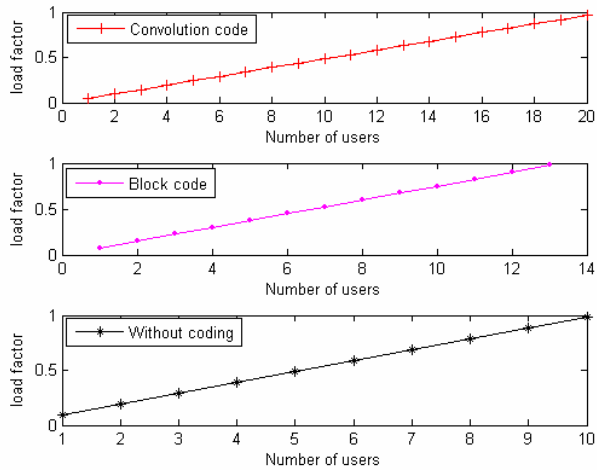

**Fig.5. Downlink load factor for video service with coding**

The Fig.5 shows the number users admitted in the forward channel of WCDMA network for video (64 kbps) service without coding scheme, with block code and convolution code. The number of users connected to the network is 7 video users for without coding scheme, 9 users for block code and 14 users for convolution code for the load factor threshold value 0.7.

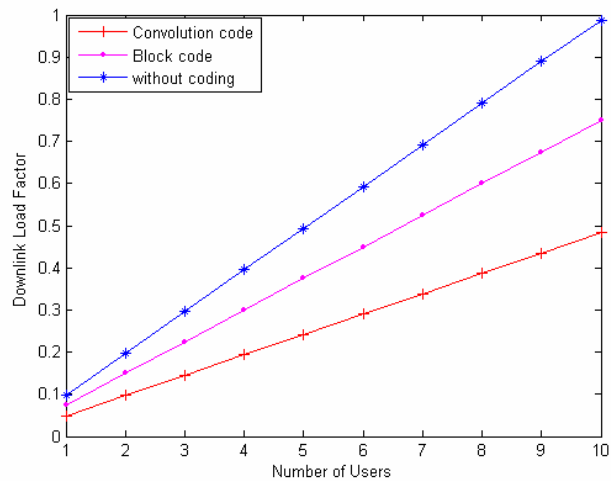

**Fig.6. Downlink load factor for video service with coding**

The Fig.6 shows the capacity enhancement of WCDMA network for video (64 kbps) service for different coding scheme. When the number of users is 10 the corresponding downlink load factor is 0.988 for without coding scheme, 0.7495 for block code and 0.4839 for convolution code

16

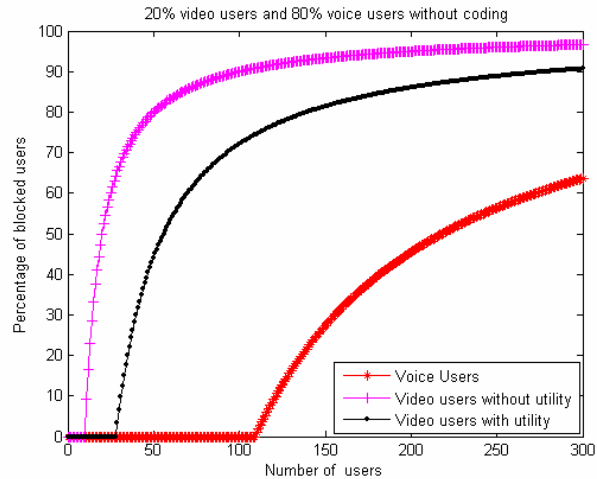

**Fig.7. Percentage of blocked users for 20% video users and 80% voice users without coding**

The Fig.7 shows the percentage of blocked users for the service utility combination of 20% video users and 80% voice users without coding scheme. The load factor threshold 0.7 in the downlink is the combination of 0.14 for video and 0.56 for voice service. When the number user is 50, the percentage of blocked user is 44% with utility function and 80% users are blocked in without utility function.

.

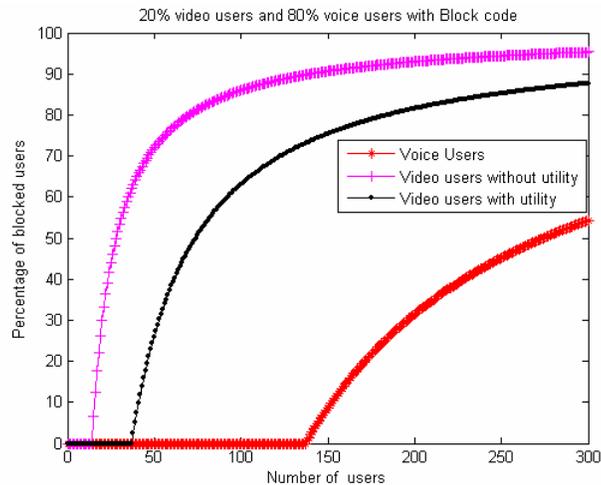

**Fig.8. Percentage of blocked users for 20% video users and 80% voice users with Block code**

The Fig.8 shows the percentage of blocked users for the service utility combination of 20% video users and 80% voice users with Block code scheme. When the number user is 50, the percentage of blocked user is 26% with utility function and 72% users are blocked in without utility function.

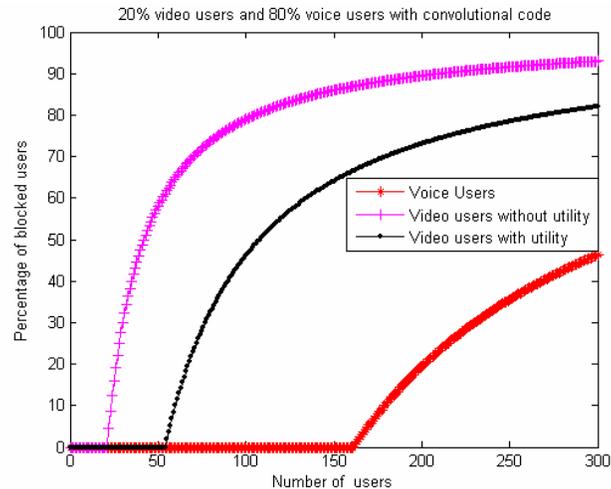

**Fig.9. Percentage of blocked users for 20% video and 80% voice users with convolutional code**

The Fig.9 shows the percentage of blocked users for the service utility combination of 20% video users and 80% voice users with convolution code scheme. When the number user is 50, the percentage of blocked user is 58% without utility function and all the users (54 users) are admitted in with utility function for 0.7 load factor.

## 4. CONCLUSION

In this paper, the main emphasis is to evaluate the performance of WCDMA network for different services such as voice (12.2 kbps) and video (64 kbps) for the downlink load factor value 0.7. The number of users admitted is calculated for different combination of services with coding scheme such as convolution code, block code and without coding for the same load factor. In the fixed threshold scheme the reserved capacity for voice service may be unutilized while video classes are blocked. In the utility based CAC scheme, the unutilized voice traffic class resources are utilized by video classes and hence reduce the blocking probability. The network capacity further enhanced with convolution code with quality of service for voice and video users by selecting appropriate value of BER.

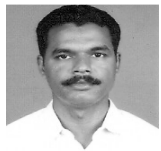
**K. Ayyappan** received the Bachelors Degree in Electronics and Communication Engineering from Bharathidasan University in 1989. He completed his Masters degree in Power Systems from Annamalai University in 1991. He is currently Professor in ECE department of Rajiv Gandhi College of Engineering and Technology, Pondicherry, India. He is pursuing research in the area of internetworking in wireless communication. He has published five papers in international journals in the same area. His areas of interest include signal processing and mobile communication.

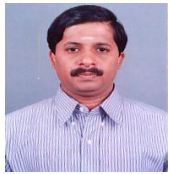
**Dr. R. Kumar** received the Bachelors Degree in Electronics and Communication Engineering from Bharathidasan University in 1989. And Master of Science in 1993 from the BITS Pilani and Ph.D. degree from SRM University, Chennai in 2009. He is working as a Professor in the Department of Electronics and Communication Engineering, SRM University, Chennai, India. He has 14 publications in National and International Journals. He is currently guiding four Ph.D students. His areas of interest include Spread spectrum Techniques and Wireless Communication.